\documentclass[prl,twocolumn]{revtex4}

\usepackage{graphicx}

\begin{document}

\title{Quantum insulating states of $F=2$ cold atoms in optical lattices}
\author{Fei Zhou and Gordon W. Semenoff}
\affiliation{ Pacific Institute of Theoretical Physics
and the Department of Physics and Astronomy, \\
The University of British Columbia, Vancouver, B. C., Cananda V6T1Z1}
\date{{\small \today}}

\begin{abstract}
In this Letter we study various spin correlated insulating states of $F=2$ cold atoms
in optical lattices.
We find that the effective spin exchange
interaction due to virtual hopping contains
an {\em octopole} coupling between two neighboring lattice sites.
Depending on scattering lengths and
numbers of particles per site
the ground states are either rotationally invariant dimer
or trimer Mott insulators or insulating states with various spin orders.
Three spin ordered insulating phases are
ferromagnetic, cyclic and nematic Mott insulators.
We estimate the phase boundaries for states with different numbers
of atoms per lattice site.
\end{abstract}
\maketitle

Recently, the scattering lengths of $Rb$ atoms in the F=2 manifold
in optical lattices have been studied
experimentally~\cite{Bloch06}. These results, together with
earlier theoretical estimates of scattering lengths~\cite{Greene}
offer critical information about how atoms interact in the $F=2$
manifold and are vital for the further understanding of their
magnetic phases. Investigation of cold atoms with high spins in
optical lattices not only improves our understanding of
fundamental ideas in quantum magnetism, but might also lead to
major breakthroughs in quantum information storing and
processing~\cite{Kitaev03,Raussendorf01}. A few impressive
theoretical efforts were already made to understand BECs of $F=2$
sodium and rubidium atoms in optical traps~\cite{Ho99,Koashi00}.
Depending on the two-particle scattering lengths in total spin
$F=0, 2, 4$ channels, the ground states of cold atoms can be
either polar, cyclic or ferromagnetic condensates. The existence
of various spin correlated states of cold atoms in optical
lattices has also been suggested~\cite{Demler02,Wu05}.

\begin{figure}[tbp]
\begin{center}
\includegraphics[width=3.5in]
{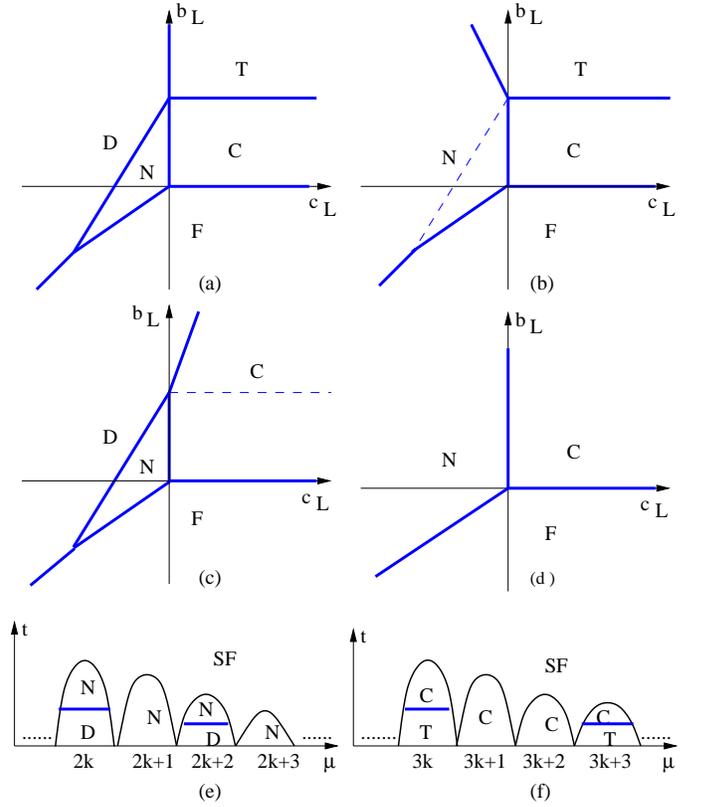}
\end{center}
\caption{ (Color online) Schematic of phase diagrams for spin
correlated Mott states of $F=2$ cold atoms in optical lattices. In
(a)-(d), the phase diagrams are plotted in a plane of two
interaction constants $b_L$, $c_L$ for different $M$ (the number
of particle per lattice site). Only insulating states are shown.
$a_L,b_L$ are defined in terms of scattering lengths after
Eq.~(\ref{Hamiltonian}). In (e) and (f), the vertical axis is the
hopping $t$ and the horizontal axis is the chemical potential
$\mu$; the numbers along the $\mu$-axis indicate the number of
particles per lattice site in Mott states (states inside lobes).
Outside the lobes are superfluids which we do not discuss in this
article. (a) for $M=3k=2k'$; (b) for $M=3k\neq2k'$ (c) for
$M=2k'\neq 3k$; (d) for $M\neq 3k\neq 2k'$ or for $M=1$. (e) for $
-c_L \gg b_L >0$; (f) for $ c_L \gg b_L >0$. N,C,F, D, T stand,
respectively, for nematic, cyclic, ferromagnetic, dimer and trimer
Mott insulating states; In (b) and (c), there are {\em no} phase
transitions across dashed lines (guide for eyes only). Phase
boundaries (thick blue lines) are estimated by applying
Eq.~(\ref{mfe}) and comparing the energies obtained in
Eq.~(\ref{energy}-\ref{oneparticle}). \label{1}}
\end{figure}

In this Letter we shall address the following question: What are
the possible insulating states of $F=2$ cold atoms in optical lattices? 
We first summarize the main results of our investigation. a) When optical
potentials are strong and exchange interactions are asymptotically
weak, for certain numbers of particles per lattice site the ground
states are rotationally invariant Mott insulators of dimers (i.e.,
singlets of two $F=2$ atoms) or trimers (i.e., singlets of three
$F=2$ atoms). b) Decreasing the optical potential depth, depending
on scattering lengths, results in various spin ordered insulating
states such as cyclic, nematic and ferromagnetic insulators. c)
The boundaries between different insulating phases are estimated
(See Fig.(\ref{1})).

To obtain these results, we consider $F=2$ cold atoms which have
two-body scattering lengths,  $a_{F=0,2,4}$, in three channels
with total hyperfine spins $F=0,2,4$. The two-body contact
interaction is $\sum_{F=0,2,4} g_F {\cal P}_F$ where $g_F=4\pi
\hbar^2 a_F/M$ and ${\cal P}_F$ are the projection operator which
projects out total spin F states of two atoms. Alternatively this
can be rewritten in terms of $a + b/2 (\hat{F}^2-12)+5 c {\cal
P}_0$ where $\hat{F}$ is the total spin operator. Three
interaction constants were calculated previously, $a= (4g_2 +
3g_4)/7$, $b=-(g_2-g_4)/7$ and $c=(g_0-g_4)/5-2(g_2-g_4)/7$
~\cite{Ho99,Koashi00}. To study the states in optical lattices, we
find it is convenient to introduce a tensor representation
$\psi^\dagger_{\alpha\beta}$, $\alpha,\beta=x,y,z$ for $F=2$
states:

\begin{eqnarray}
&& \psi^\dagger_{xz}=\frac{1}{\sqrt{2}} (\psi^\dagger_{-1} -
\psi^\dagger_1)~~,~~ \psi^\dagger_{yz}=\frac{i}{\sqrt{2}}
(\psi^\dagger_{-1}+ \psi^\dagger_1),
\nonumber \\
&&
\psi^\dagger_{xy}=\frac{i}{\sqrt{2}}(\psi^\dagger_{-2}-\psi^\dagger_2)~~,~~
\psi^\dagger_{zz}=\frac{2}{\sqrt{3}}
\psi^\dagger_0, \nonumber\\
&& \psi^\dagger_{xx}=\frac{
1}{\sqrt{2}}(\psi^\dagger_{-2}+\psi^\dagger_{2})-\frac{1}{\sqrt{3}}\psi^\dagger_0
\nonumber\\
&& \psi^\dagger_{yy}=\frac{-
1}{\sqrt{2}}(\psi^\dagger_{-2}+\psi^\dagger_{2})-\frac{1}{\sqrt{3}}\psi^\dagger_0
\end{eqnarray}
where $\psi^\dagger_{m_F}$, $m_F=0,\pm 1, \pm 2$ are the usual
creation operators for $F=2$ particles. The tensor operators
$\psi^\dagger_{\alpha\beta}$ are symmetric and traceless, that is
$\psi^\dagger_{\alpha\beta}=\psi^\dagger_{\beta\alpha}$ and $tr
\psi^\dagger=0$. We find this a convenient labelling of the five
$F=2$ states. The commutation relations are

\begin{eqnarray}
[\psi_{\alpha\beta},\psi^\dagger_{\alpha'\beta'}]=
\delta_{\alpha\alpha'}\delta_{\beta\beta'}+
\delta_{\alpha\beta'}\delta_{\beta\alpha'}
-\frac{2}{3}
\delta_{\alpha\beta}\delta_{\alpha'\beta'}.
\end{eqnarray}

The construction of rotationally invariant operators in this
representation is straightforward. For instance, the number
operator $\hat{\rho}$, the dimer or singlet pair creation operator
${\cal D}^\dagger$ and the trimer or singlet of three atoms
creation operator ${\cal T}^\dagger$ are
\begin{equation}
\hat{\rho}=\frac{1}{2}tr \psi^\dagger \psi,~ {\cal
D}^\dagger=\frac{1}{\sqrt{40}}tr \psi^\dagger\psi^\dagger,~ {\cal
T}^\dagger=\frac{1}{\sqrt{140}}tr
\psi^\dagger\psi^\dagger\psi^\dagger
\end{equation}
where $\psi^\dagger$ represents the tensor.
The total spin operator $\hat{F}_\alpha$, $\alpha=x,y,z$ is

\begin{equation}
\hat{F}_\alpha=-i\epsilon_{\alpha\beta\gamma}
\psi^\dagger_{\beta\eta}\psi_{\eta\gamma},~~
[\hat{F}_\alpha,\hat{F}_\beta]=i\epsilon_{\alpha\beta\gamma}\hat{F}_\gamma.
\end{equation}
Finally, $\hat{F}_\alpha$ commutes with rotationally invariant operators such as
${\cal D}^\dagger$ and ${\cal T}^\dagger$; generally one finds that
$[\hat{F}_\alpha, tr (\psi^\dagger)^n]=0$.
For $F=2$ cold atoms in optical lattices, we employ the following Hamiltonian

\begin{eqnarray}
&& {\cal H}=\frac{a_L}{2} \sum_{k} (\hat{\rho}^2_k-\hat{\rho}_k)
+\frac{b_L}{2} \sum_{k}\big(\hat{F}^2_k - 6\hat{\rho}_k \big) \nonumber \\
&& + 5 c_L \sum_k {\cal D}^\dagger_k {\cal D}_k
-t\sum_{<kl>} (\psi^\dagger_{k,\alpha\beta} \psi_{l,\beta\alpha} + h.c.).
\label{Hamiltonian}
\end{eqnarray}
Here $k$ is a lattice site index and $<kl>$ are neighboring sites.
$t$ is the one-particle hopping amplitude. $a_L, b_L, c_L$
are, respectively, proportional to $a,
b,c$ introduced earlier; $a_L(b_L,c_L) =a(b,c)\int d{\bf r} (\psi_0^*\psi_0)^2$ and $\psi_0$
is the on-site local ground state wavefunction.
${\cal D}^\dagger{\cal D}$ plays the role of projection operator ${\cal P}_0$
discussed before.
For most cold atoms, the absolute values of $b_L$, $c_L$ are much
smaller than $a_L$ and we work in this limit unless specified.
When the optical potential depth is small and $t$ is much larger than $a_L$,
cold atoms effectively are weakly interacting and condense.
The results obtained using Eq.(\ref{Hamiltonian}) are identical to the ones previously obtained and
we do not present them in this Letter. Here we focus on the physics of
insulating states when $t$ is much smaller than $a_L$.

The effective low energy Hamiltonian for a Mott insulator with $M(>1)$ atoms per lattice site is

\begin{eqnarray}
&& {\cal H}=
\frac{b_L}{2} \sum_{k} \big( \hat{F}^2_k - 6\hat{\rho}_k \big) + 5 c_L \sum_k {\cal D}^\dagger_k
{\cal D}_k \nonumber \\
&& - J_{0} \sum_{<kl>} \big( Q^\dagger_{k;\alpha\beta,\alpha'\beta'}
Q_{l;\alpha\beta,\alpha'\beta'} + h.c.\big);
\nonumber \\
&&
Q^\dagger_{k;\alpha\beta,\alpha'\beta'}=\psi^\dagger_{k,\alpha\beta}\psi_{k,\alpha'\beta'}-
\nonumber \\
&& \frac{1}{10}tr \psi^\dagger_k \psi_k
\big(\delta_{\alpha'\alpha}\delta_{\beta\beta'}+\delta_{\alpha\beta'}\delta_{\beta\alpha'}
-\frac{2}{3}\delta_{\alpha\beta}\delta_{\alpha'\beta'}\big)
\end{eqnarray}
where $J_{0}= t^2/a_L$. The virtual hopping between two
neighboring sites results in an exchange interaction of the form
of {\em octopole} coupling as shown above. This is valid for all
Mott states with the number of particles $M$ larger than one. An
exchange interaction of quadrupole type has been derived recently
for $F=1$ cold atoms in optical
lattices~\cite{Demler02,Imambekov03,Zhou03}.

The spin-ordered Mott insulating states under consideration (with the crystal
translational symmetry) are

\begin{eqnarray}
|\chi>=\prod_{k} \frac{(\psi_k^\dagger\chi)^M}{\sqrt{M!}}|0>, tr \chi
\chi^*=\frac{1}{2};
\label{trialwf1}
\end{eqnarray}
$\chi$ is a traceless symmetric complex tensor
and $M$ is the number of atoms per site.
We calculate
the $\chi$-dependent part of the mean field energy per particle,
\begin{eqnarray}
&& E_{MI}=(M-1) A({\chi,\chi^*}) -\frac{16}{5} J_{ex} M; \nonumber \\
&& A(\chi,\chi^*)=4b_L Tr[\chi, \chi^*]^2 + 2c_L tr \chi^2 Tr
{\chi^*}^2 \big. \label{mfe}
\end{eqnarray}
Here $J_{ex}=zJ_0$, $z$ is the number of neighboring lattice sites.
Minimization of this energy subject to the constraint of the normalization of $\chi$ in
Eq.(\ref{trialwf1}) can be carried out using the method of the Lagrangian multiplier.
We provide the results here:

1) $b_L >0$ and $c_L>0$. The traceless tensor
$\chi$ satisfies
\begin{eqnarray}
tr \chi^2=[\chi,\chi^*]=0.
\end{eqnarray}
There are two discrete {\em root} solutions $\chi_{R\pm}$ which have only
diagonal components,
$\chi_{xx}=1/\sqrt{6}$, $\chi_{yy}=\exp(\pm i 2\pi/3)/\sqrt{6}$ and
$\chi_{zz}=\exp(\pm i4\pi/3)/\sqrt{6}$ and
$\chi_{\alpha\beta}=0$ when $\alpha\neq \beta$.
They are also invariant under a cyclic subgroup $C_3$ of the SO(3) rotation group.
The full set of solutions can be obtained by applying SO(3) rotation,
and $U(1)$ gauge transformation.
These solutions correspond to cyclic insulating states.

2) $b_L > {c_L}/{4}$ and $c_L < 0$. Up to a phase, $\chi$ satisfies
\begin{eqnarray}
tr \chi^2=\frac{1}{2}, \chi=\chi^*.
\end{eqnarray}
For diagonal matrices which satisfy the above conditions, the matrix elements have to fall on an
elliptical curve;
that is,
$\chi^2_{xx}+$ $\chi_{xx}\chi_{yy}$ $+\chi^2_{yy}=1/4$, $tr \chi=0$
and $\chi_{\alpha\beta}=0$ if $\alpha\neq \beta$.
The full set of solutions can be obtained by applying $SO(3)$ and $U(1)$ rotation.
These states are characterized by {\em traceless real symmetric} tensors
(up to a phase) and $<\hat{F}_\alpha>=0$, $\alpha=x,y,z$; moreover,
the corresponding expectation value of the nematic order operator $\psi^\dagger_{\alpha\gamma}\psi_{\gamma\beta}-1/3
\delta_{\alpha\beta} tr \psi^\dagger\psi$ is nonzero.
They represent nematic insulating states
(see more discussions in Ref.\onlinecite{Zhou03}).

3) $b_L <0$ and $b_L < {c_L}/{4}$. $\chi$ satisfies
\begin{eqnarray}
tr\chi^2=0,
\chi^* \chi \chi^*-\chi^*\chi^*\chi -tr [\chi,\chi^*]^2 \chi^*=0
\end{eqnarray}
and $[\chi,\chi^*]\neq 0$.
An example of solutions is $\chi_{z\alpha}=0$($\alpha=x,y,z$) and $\chi_{xx}=$
$-\chi_{yy}$ $=-i\chi_{xy}=1/\sqrt{8}$.
This corresponds to a ferromagnetic insulating phase.

So three spin ordered Mott insulating states are
ferromagnetic, nematic and cyclic states (indicated by subscripts $F, N, C$ respectively);
the corresponding energies per atom are
\begin{eqnarray}
&& E_F=2(M-1)b_L -\frac{16}{5} MJ_{ex} \nonumber \\
&& E_N=(M-1)\frac{c_L}{2} -\frac{16}{5}M J_{ex}, E_C=-\frac{16}{5} M J_{ex}.
\label{energy}
\end{eqnarray}
In addition, one should also take into account rotationally invariant
insulating states.
The physics of these states depends on the numbers of
particles per
lattice site and below we discuss different situations separately.

i)$M=2k=3k'$ ($k,k'$ are integers).
When $b_L$ and $-c_L$ are positive and $J_{ex}$ approaches zero,
the Hamiltonian has the
lowest energy eigenstate
which is rotationally invariant and has a maximal number of singlet pairs,
or {\em dimers}. It has the following wavefunction

\begin{eqnarray}
|D> =\prod_k\big( \frac{tr
\psi_k^\dagger\psi_k^\dagger}{\sqrt{40}}\big)^{\frac{M}{2}}|0>,
E_D=(M+3)\frac{c_L}{2} -3 b_L. \label{dimer}
\end{eqnarray}
The state $|D>$ (up to a normalization factor) is an insulator of singlet pairs or dimers;
following Eq.(\ref{energy}) and Eq.(\ref{dimer}),
it has a lower energy than the nematic state when the exchange
interaction is small, i.e.,
\begin{equation}
J_{ex}< J^c_{D},  J^c_D= \frac{5}{16M}\big(3b_L-2c_L \big).
\end{equation}
In dimer Mott states, all atoms are paired in singlets and $<D|{\cal
D}^\dagger{\cal D}|D>=M(M+3)/10$.

When $b_L$ and $c_L$ are both positive and the exchange coupling is zero,
the ground state of the Hamiltonian
is still an M-atom spin singlet but has to be {\em annihilated} by the
dimer operator ${\cal D}$.
For $M=3$, such a state can be created by the trimer creation operator ${\cal T}$;
the wavefunction and its energy are

\begin{eqnarray}
|T> =\prod_k \frac{1}{\sqrt{140}}tr (\psi_k^\dagger)^3|0>, E_T=-3 b_L
\label{trimer}
\end{eqnarray}
Note that ${\cal D}|T>=0$ and $<T|{\cal D}^\dagger{\cal D}|T>=0$.

For more than three particles, one can show that singlet states of
the form $tr\psi^n$ with $n>3$ can always be rewritten in terms of
two {\em fundamental} singlet creation operators: $tr\psi^2$ for a
dimer and $tr \psi^3$ for a trimer. Generally, singlet states for
$M=2k_1+3k_2$ atoms can be expressed in terms of $(tr
\psi^\dagger\psi^\dagger)^{k_1}$
$(tr\psi^\dagger\psi^\dagger\psi^\dagger)^{k_2}$. For instance,
direct calculations show that $tr (\psi^\dagger)^4|0>$ is
identical to $(tr \psi^\dagger\psi^\dagger)^2|0>$ and
$tr(\psi^\dagger)^5|0>$ is identical to $tr (\psi^\dagger)^2 tr
(\psi^\dagger)^3|0>$. To construct an M-particle singlet without
singlet pairs, the only candidate to consider is
$(tr\psi^\dagger\psi^\dagger\psi^\dagger)^k|0>$ since all other
states explicitly involve $(tr \psi^\dagger\psi^\dagger)^n$, the
dimer creation operator. Therefore, for an arbitrary M, a 3k-atom
singlet {\em without dimers} can be written as $|T>= \prod_k {\cal
P}_T(k) (tr (\psi^\dagger)^3)^k$; (up to the normalization factor)
here ${\cal P}_T(k)$ is a projection operator which further
projects out states without dimers at site $k$, i.e. ${\cal D}_k
{\cal P}_T(k)=0$. Taking into account that $\hat{F}_k^2|T>={\cal
D}_k|T>=0$, we find that the energy per atom for a trimer state is
independent of $M$, $E_T=- 3 b_L$. It has a lower energy than the
cyclic state when

\begin{equation}
J_{ex} < J^c_{T}, J^c_T= \frac{15}{16M} b_L.
\end{equation}
By further comparing $E_{D,T}$ and $E_{C, N, F}$ we obtain the mean field phase
diagram;
we find the mean field phase boundaries between rotationally invariant
trimer and dimer states and
three spin ordered phases, taking into account Eq.(\ref{energy}), Eq.({\ref{dimer})
and Eq.({\ref{trimer}).
The results are summarized in Fig.(\ref{1} a).
The boundary between $T$- and $D$-phase is $c_L=0$, $C$- and $F$-phase is
$b_L=0$,
$T$ and $C$-phase is $b_L=16 M J_{ex}/15$,
between $D$- and $N$-phase is
$b_L=2 c_L/3 +16 M J_{ex}/15$, between $N$- and $F$-phase
is $b_L=c_L /4$, and finally between $D$ and $F$-phase is
$b_L=(M+3)c_L/(4M+2)
+16 M J_{ex}/5(2M+1)$.

ii) $M=3k= 2k'+ 1$. Following the general construction outlined in
i), the trimer states are ground states in optical lattices when
$c_L$ and $b_L$ are both positive and $J_{ex}$ is approaching
zero. However, when $c_L$ is negative, $b_L$ is positive, as
$J_{ex}$ approaches zero, at each lattice site $k'$ singlet pairs
or dimers are formed leaving the last atom unpaired. So the system
effectively is a lattice with an $S=2$ spin at each lattice site.
The situation is similar to what happens to $F=1$ atoms in optical
lattices with an odd number of atoms per site~\cite{Zhou03}.
Following the discussions on the case for one atom per lattice
($M=1$) below, we find that the ground state remains to be
nematically ordered down to $J_{ex}=0$ as far as $c_L$ is negative
and $b_L$ is positive. Only four phases (trimer Mott insulators
and spin ordered cyclic, ferromagnetic, nematic Mott states)
appear in the phase diagram as shown in Fig.(\ref{1} b).
%Along the dashed line $b_L=2 c_L/3 +16 M J_{ex}/15$
%there are no phase transitions.
The phase boundary between
$T$- and $N$-phase is $b_L=(1-M)c_L/6+16 M J_{ex}/15$ which approaches
$c_L=0$ as $M$ becomes infinity.
The other phase boundaries
are the same as in Fig. (\ref{1} b) discussed before.

iii) $M=2k=3k'\pm 1$. When $c_L$ is negative and $b_L$ is positive
and the exchange interactions are small, the ground states are
dimer insulating states. When $c_L$ and $b_L$ are positive and
$J_{ex}$ approaches zero, at each lattice site $k'$ trimers are
formed; the last atom (for $M=3k'+1$) or two (for $M=3k'+2$) which
do not participate in the trimer formation are in an $F=2$ spin
collective state. Effectively at each site only the $F=2$ state is
relevant for spin ordering and the system is again equivalent to a
lattice spin Model for $S=2$ spins. This limit is equivalent to
$M=1$ case studied below and we find that the cyclic state remains
to be the ground state as $J_{ex}$ becomes zero, provided that
$b_L$ and $c_L$ are positive. Four phases that appear in the phase
diagram are dimer Mott insulators and spin ordered cyclic,
ferromagnetic, nematic Mott states, as shown in Fig.(\ref{1} c).
The boundary between $D$ and $C$ is $b_L= 16 M J_{ex}/15 + (M+3)/6
c_L$; again it approaches $c_L=0$ as $M$ becomes infinity. The
other boundaries are identical to those in Fig.(\ref{1} a).

iv)
Finally, we would like to address the special case of $M=1$.
The effective Hamiltonian derived before
is only valid when $M>1$. Here
we directly calculate the exchange interaction between two
adjacent $F=2$ spins due to virtual hopping and obtain the following
Hamiltonian
\begin{eqnarray}
\frac{\cal H}{J_{0}}=-\sum_{<kl>} \sum_{F=0,2,4} \frac{\tilde{a}}{{a}_F}
{\cal P}_{F}(kl)
\label{Hamiltonian1p}
\end{eqnarray}
Here ${\cal P}_F(kl)$ projects out a total spin $F$ state of two
atoms on neighboring sites $<kl>$.  $\tilde{a}$ is defined as
$\tilde{a}=(4a_2+3a_4)/7$. For two adjacent atoms with total spin
$F=1,3$, the exchange interaction vanishes. We calculate the
energy per particle using the wavefunction specified by
Eq.(\ref{trialwf1}) with $M=1$ and derive an expression for the
energy as a function of $\chi$. The $\chi$-dependence of the energy per atom
can be shown identical to that in Eq.(\ref{mfe}).
We again obtain three phases and the
energies per atom of cyclic, nematic and ferromagnetic states are

\begin{eqnarray}
&& \frac{E_F}{J_0}=-\frac{\tilde{a}}{a_4},
\frac{E_c}{J_0}=-\frac{3\tilde{a}}{7a_4}
-\frac{4\tilde{a}}{7a_2}, \nonumber \\
&& \frac{E_N}{J_0}=-\frac{18\tilde{a}}{35a_4} -
\frac{2\tilde{a}}{7a_2}-\frac{\tilde{a}}{5a_0}.
\label{oneparticle}
\end{eqnarray}
When the difference between scattering lengths is small, one can
easily verify that the phase boundaries of these three insulating
states are identical to those of three condensates discussed
previously. The results are summarized in Fig.(\ref{1} d). The
boundaries between $F$ and $C$, $C$ and $N$ and $N$ and $F$ are,
respectively, $b_L=0$, $c_L=0$ and $b_L=c_L/4$~\cite{Zawitkowski06}.
Finally we present the results in terms of hopping $t$ and
chemical potential $\mu$ of $F=2$ atoms in two parameter regions: 1)
$-c_L \gg b_L >0$  ( Fig.(\ref{1}e)); and 2) $c_L \gg b_L >0$ 
(Fig.(\ref{1}f) ); the spin ordering exhibits mod 2
(Fig.(\ref{1}e)) and mod 3 (Fig.\ref{1}f) behaviors.

In conclusion, we find that $F=2$ cold atoms in optical lattices in
addition to forming spin ordered Mott states, can also be in states
that only consist of dimers or trimers. These results 
are relevant to $Rb$ and $Na$ atoms in
optical lattices. According to earlier estimates\cite{Greene},
$^{87}Rb$ atoms fall nearly on the $c_L = 0$ line ($b_L>0$), a phase 
boundary between the cyclic and nematic insulating phases. For $^{23}Na$, 
$b_L = -4.61 c_L >0$; the insulating state of $^{23}Na$ atoms is nematic when
$J_{ex}$ is large; for $M=2k$ as the optical potential decreases,
a quantum phase transition occurs and
the ground state becomes
a dimer Mott state. 
However, the error bars in estimates of scattering lengths might lead to 
uncertainties in determining ground states. 
Finally, dimer or trimer Mott states have fully gapped spin excitations and are
robust; spin ordered states in $3D$ optical lattices are stable 
against quantum fluctuations (see similar discussions in Ref.\cite{Zhou03}).
In a finite trap because of coexistence of Mott states with different occupation
numbers\cite{Jaksch98}, the spin order spatially alternates between dimer 
(trimer) and nematic (trimer) ones. This work is 
supported by the office of the Dean of Science, UBC, and
NSERC, Canada. FZ currently is an A. P. Sloan fellow.

\end{document}